\documentclass[reprint,twocolumn,aps,prb,10pt,amsmath,amssymb,longbibliography,superscriptaddress, showpacs,preprintnumbers,nofootinbib]{revtex4-2}

\usepackage{graphicx}%
\usepackage{siunitx}
\usepackage{bm}
\usepackage[caption=false]{subfig}
\usepackage{mathtools}
\usepackage{multirow}
\usepackage{amssymb}
\usepackage{braket}
\usepackage{amsmath}
\usepackage{placeins}
\usepackage{cancel}
\usepackage{xcolor}
\usepackage{physics}
\usepackage{spreadtab}
\usepackage{dsfont}
\usepackage[normalem]{ulem}  
\usepackage[colorlinks=true,
            linkcolor=blue,
	        citecolor=blue,
	         urlcolor=blue]{hyperref}

\begin{document}

\title{Challenging the $p$-type Paradigm: Intrinsic $n$-type Mobility in Antiferromagnetic Cr$_2$O$_3$}

\author{\'Alvaro Adri\'an Carrasco \'Alvarez}
\email{alvaro.carrasco@uclouvain.be}
\affiliation{European Theoretical Spectroscopy Facility, Institute of Condensed Matter and Nanosciences, Université catholique de Louvain, Chemin des Étoiles 8, B-1348 Louvain-la-Neuve, Belgium}
\author{Samuel Ponc\'e}%
\email{samuel.ponce@uclouvain.be}
\affiliation{European Theoretical Spectroscopy Facility, Institute of Condensed Matter and Nanosciences, Université catholique de Louvain, Chemin des Étoiles 8, B-1348 Louvain-la-Neuve, Belgium}
\affiliation{WEL Research Institute, avenue Pasteur 6, 1300 Wavre, Belgium.}

\date{\today}

\begin{abstract}
Chromium oxide (Cr$_2$O$_3$) is widely considered a $p$-type transparent conducting oxide despite ongoing debate regarding its intrinsic transport character. Here, we resolve this question by computing phonon-limited electron and hole mobilities using the \textit{ab initio} Boltzmann transport equation. We find that electron mobility systematically exceeds hole mobility over a wide temperature range, demonstrating that Cr$_2$O$_3$ is intrinsically $n$-type.
Analysis of scattering mechanisms reveals that scattering with phonons affects electrons and holes similarly, and that the mobility asymmetry originates from the electronic structure, namely the larger effective mass and multi-valley character of the valence band.
The intrinsic $n$-type character, combined with moderate hole mobility, enables bipolar transport and revises the role of Cr$_2$O$_3$ in transparent electronics. Additionally, our results on mobility complement previous studies on defect formation indicating that the commonly observed $p$-type behavior is extrinsic. These insights provide a complete chemical-transport paradigm for Cr$_2$O$_3$, re-evaluating its role in functional transparent electronic and magneto-optoelectronic applications
\end{abstract}

\maketitle

\section{Introduction} Transparent conducting oxides (TCO) have attracted interest in the last decade due to their potential in optoelectronic applications~\cite{Zhang2016,Morales2017,Dixon2016,Rebecchi2023}. 
These types of materials offer good electrical conductivity and are transparent in the visible range, making them interesting candidates for use on screens in mobile phones, tablets, or touchscreens~\cite{Stadler2012,Minami2000}.
The most commonly used materials are $n$-type semiconductors offering high conductivity and transparency, and a more stable doping~\cite{Zhang2019,Cui2023,Johnson2021,Hu2019}. 
As a consequence, this means that in most practical applications of TCOs, $n$-type semiconductors are preferred. 
Nonetheless, having a $p$-type TCO is crucial for making $p-n$ junctions for transparent light emitting diodes~\cite{Kawazoe2000,Nagarajan2001}, transparent photovoltaics~\cite{Jang2019,Moreira2022,Fioretti2020,Patel2023} or to obtain transparent photo-detectors, logic devices and complementary metal oxide semiconductors (CMOS) \cite{Ohta2004,Ozel2024,Rana2019}. 
Compounds such as Cr$_2$O$_3$ have  been proposed to be good candidates for $p$-type doping \cite{Kehoe2016,Dabaghmanesh2017}, even suggesting Cr$_2$O$_3$ to be an intrinsic $p$-type semiconductor \cite{Iordanova2005,Crawford1964}. 

In that regard, several experimental groups have been successful in this pursuit, synthesizing $p$-type Cr$_2$O$_3$ samples~\cite{Mohanapandian2021,Arca2017,Uekawa1996,Arca2011,Jagadish2024,Park1990,Jella2017,Crawford1964}. 
Nonetheless, it has been reported that $p$-type doping of Cr$_2$O$_3$ is challenging~\cite{Uekawa1996,Arca2011,Jagadish2024} and that $n$-type doping is also possible~\cite{Jella2017,Holt1999}, as well as some conflicting results with respect to oxygen deficiency and fabrication process~\cite{Holt1999,Su1989,Parsa2018,Kofstad1980}. 
Surprisingly, there are few theoretical studies~\cite{Iordanova2005,Dabaghmanesh2017} that address the conducting properties of Cr$_2$O$_3$. 
Thus, the question of the  intrinsic nature of Cr$_2$O$_3$ remains open.
\begin{table}[b]
    \centering
    \begin{tabular}{l l l l l}
    \hline\hline
        Material & $\mu^\mathrm{e}$ (cm$^2$/(V$\cdot$s)) & $\mu^\mathrm{h}$ (cm$^2$/(V$\cdot$s))  &  $\mu^\mathrm{f}/\mu^\mathrm{s}$ & E$_g$ (eV) \\ \hline
        In$_2$O$_3$    & 50-100~\cite{Newhouse2005} & 0.1$<$~\cite{Ou2008} & 500$>$ & 2.7~\cite{deBoer2016,Nagata2017} \\
        SnO$_2$    & 20-200~\cite{Li2022}   & 1-4~\cite{Li2022} & 5-50 & 3.7~\cite{Musztyfaga2020} \\
        ITO & 50-100~\cite{Zlotnik2022}  & 0.1$<$~\cite{Mun2025} & 500$>$ & 4.0~\cite{Musztyfaga2020} \\
        ZnO &  10-200~\cite{Hutson1959} & 0.1$<$~\cite{Mora2020,Hammer2010}  & 100$>$  & 3.4~\cite{Ahmad2025}  \\
        Ga$_2$O$_3$ & 100-200~\cite{Zhang2025} & 1-5~\cite{Ma2022} & 40-100   & 4.6-4.8~\cite{Peelaers2017} \\
        IGZO & 20-50~\cite{Shin2017} & - & - & 3.0-3.2~\cite{Hays2017} \\
        CuAlO$_2$ & - & 0.1-1~\cite{Li2018,Yao2012} & - & 3.5~\cite{Pellicer2006} \\
        SnO    & 0.2~\cite{Mun2025}  & 1-7~\cite{Chae2023}   &  5-15 & 2.7~\cite{Ogo2008} \\[1ex]        
        \multicolumn{5}{c}{This work} \\
        \hline 
        Cr$_2$O$_3$ & 5-10 & 1-4 & 2-3 & 3.1~\cite{Patel2023} \\ \hline\hline
    \end{tabular}
    \caption{Electrical and optical properties, Band gap E$_g$, electron (hole) mobilities $\mu^\mathrm{e}(\mu^\mathrm{h})$ and ratio $\mu^\mathrm{f}/\mu^\mathrm{s}$ between the fastest $\mu^\mathrm{f}$ and slowest $\mu^\mathrm{s}$ carrier type mobility, of common TCO materials  compared with Cr$_2$O$_3$, where ITO is In$_2$O$_3$ and SnO$_2$, and IGZO  is In$_2$O$_3$, Ga$_2$O$_3$ and ZnO.}
    \label{TCO-table}
\end{table}

In this study, we use spin polarized Density Functional Theory (DFT) calculations and the Boltzmann transport equation (BTE)~\cite{Ponce2020,Claes2025,Carrascolvarez2026}, a methodology previously unexplored for this system, to compute the electron and hole mobilities of bulk antiferromagnetic Cr$_2$O$_3$.
We verify its applicability in Supplementary Material S1~\cite{MCA2026}. 
We find that (i) the electron mobility of Cr$_2$O$_3$ is consistently higher than the hole mobility in the 100-500~K temperature range, implying that Cr$_2$O$_3$ is an intrinsic $n$-type semiconductor; 
(ii) the hole mobility ($\mu^\mathrm{h}$) of Cr$_2$O$_3$ is between $1-4$~$\mathrm{cm^2/(V\cdot s)}$ at room temperature, which in the context of $p$-type TCOs makes it competitive with already established materials; and (iii) the conducting easy axis aligns with the magnetic one.

These findings provide an unexpected perspective on Cr$_2$O$_3$, shifting its description from a strictly $p$-type compound to an intrinsic $n$-type wide band gap semiconductor. Crucially, as we show in Table~\ref{TCO-table}, Cr$_2$O$_3$ exhibits a near unity electron-hole mobility ratio $\mu^\mathrm{e}/\mu^\mathrm{h} = 2-3$ absent in all high-performance $n$-type TCOs. This intrinsic transport parity provides a unique opportunity for balanced charge extraction in all-oxide bipolar architectures. By establishing a solid link between electronic transport properties and the underlying thermodynamic defect landscape, this work outlines targeted synthesis routes to exploit the true, intrinsic mobile carrier channels of this stable oxide. Furthermore, the alignment of the magnetic and conducting easy axes, provides a unique platform for magneto-optoelectronic tuning, bridging the gap between transparent circuitry and antiferromagnetic spintronics. 

\section{Methods} 
\begin{figure*}[t]
    \centering
    \includegraphics[width=0.7\linewidth]{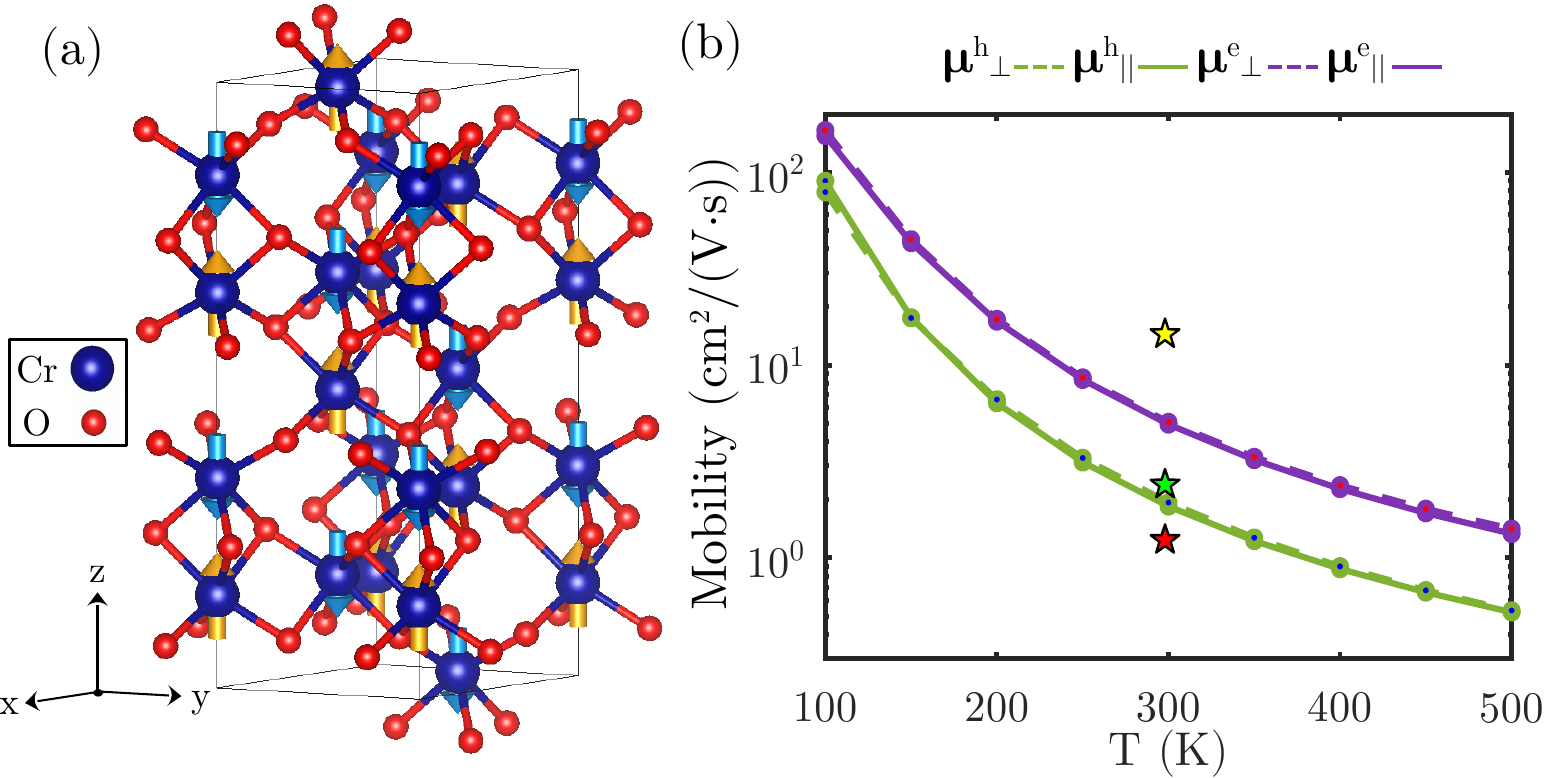}
    \caption{\label{Cr2O3-mob}
    (a) Crystal structure of antiferromagnetic Cr$_2$O$_3$ with the arrows depicting the collinear magnetic order. 
    (b) Intrinsic electron (in-plane $\mu^{\rm e}_{\parallel}$, dash mauve, and out-of-plane $\mu^{\rm e}_{\perp}$, mauve line) and hole (in-plane $\mu^{\rm h}_{\parallel}$, dash green, and out-of-plane $\mu^{\rm h}_{\perp}$, green line) mobility for bulk polycrystalline Cr$_2$O$_3$ as a function of temperature obtained with the Boltzmann transport equation. 
    Solid star symbols correspond to the experimental drift mobility from Refs.~\cite{Singh2025,Qiu2024} at room temperature.}
\end{figure*}

We use DFT with the \textsc{Quantum ESPRESSO} v7.6 software~\cite{Giannozzi2009,Giannozzi2017}.
Scalar-relativistic norm-conserving pseudopotentials~\cite{Hamann2013} from the standard accuracy table of the \textsc{PseudoDojo} project~\cite{vanSetten2018} are used with the Pardew Burke Ernzerhof (PBE) functional~\cite{Perdew1996}.
The plane-wave cutoff is set to 90~Ry, a total energy convergence threshold of $10^{-12}$~Ry, a $4^3$ $\mathbf{k}$-point grid, and a $6^3$ $\mathbf{q}$-point grid for phonons.
We relax the crystal structure in its $R\overline{3}c$ ground state geometry with the N\'eel antiferromagnetic order (AFM) as shown in Fig.~\ref{Cr2O3-mob}(a),  
obtaining $a = 4.948~\mathrm{\AA}$ and $c = 13.781~\mathrm{\AA}$ lattice parameters, close to the experimental values of $a_\mathrm{exp} = 4.953~\mathrm{\AA}$ and $c_\mathrm{exp} = 13.578~\mathrm{\AA}$~\cite{Abdullah2014}.
The phonons are computed with density functional perturbation theory (DFPT)~\cite{Baroni1987,Savrasov1992,Gonze1997,Baroni2001}. 
The maximally localized Wannier functions are computed using the \textsc{wannier90} package version 3.1~\cite{Pizzi2020} with Cr $d$ and O $p$ initial projections. 
The phonon limited transport is computed with the  \textsc{EPW} package~\cite{Lee2023,Yang2025,Carrascolvarez2026} which will be released in the v6.1. Mobilities are converged within 1\% using $6^3$ coarse and $32^3$ fine $\mathbf{k}/\mathbf{q}$ grids.
We provide all data to reproduce the results on the Materials Cloud Archive~\cite{MCA2026}.

The intrinsic electron and hole mobility of Cr$_2$O$_3$ are calculated by means of the iterative \textit{ab initio} BTE~\cite{Ponce2018,Claes2025} in the range of 100-500~K. Although Cr$_2$O$_3$ presents a N\'eel temperature of $T_{\text{N\'eel}} = 307$~K~\cite{Gurevich2009,Anderson1937}, there are no significant changes in the experimental conductivity that may signal a different trend~\cite{Mohanapandian2021,Arca2017,Uekawa1996,Arca2011,Jagadish2024,Park1990,Jella2017,Crawford1964}.  
Consequently, the given trends in mobility on the AFM phase would still hold in the paramagnetic phase. 
We consider that each spin channel $\sigma$ contributes with a spin resolved mobility tensor $\mu_{\alpha\beta}^\sigma$ defined as~\cite{Carrascolvarez2026} 
\begin{equation}
    \mu_{\alpha\beta}^{\sigma} = \frac{-e}{V_{\rm uc}n_c^\sigma}\sum_{n}\int\frac{\mathrm{d}^3\mathbf{k}}{\Omega_\mathrm{BZ}}v^{\sigma}_{n\mathbf{k}\alpha} \partial_{E_\beta} f_{n\mathbf{k}}^\sigma,
\end{equation}
where $V_{\rm uc}$ is the volume of the unit cell, $\Omega_\mathrm{BZ}$ is the volume of the Brillouin Zone, $\partial_{E_\beta} f_{n\mathbf{k}}^\sigma$ is the perturbed occupation for an applied electric field $E_\beta$ in the direction $\beta$, $v_{n\mathbf{k}\alpha}^{\sigma}$ is the $\alpha$ component of the spin-resolved electronic band velocity and $n_c^\sigma$ is the carrier density of a specific spin channel such that the relation with the conductivity tensor $\sigma_{\alpha\beta}$ is
\begin{equation}
    \sigma_{\alpha\beta} = n_c^\uparrow\mu_{\alpha\beta}^\uparrow +  n_c^\downarrow\mu_{\alpha\beta}^\downarrow.
\end{equation}
We note that the directly measurable quantities are the total conductivity $\sigma_{\alpha\beta}$ and its related mobility defined as
\begin{equation}
    \mu^\mathrm{eff}_{\alpha\beta}  \equiv\frac{\sigma_{\alpha\beta}}{n_c} = \sum_\sigma\frac{n_c^\sigma}{n_c}\mu^\sigma_{\alpha\beta},
\end{equation}
where $n_c \equiv n_c^\uparrow+n_c^\downarrow$.
In an antiferromagnet such as Cr$_2$O$_3$ where the bands are spin degenerate, the carrier density on each spin channel is the same $n_c^\uparrow = n_c^\downarrow$. 
In practice, this means that the effective mobility is $\mu^\mathrm{eff}_{\alpha\beta} = (\mu^\uparrow_{\alpha\beta}+\mu^\downarrow_{\alpha\beta})/2$. 
In addition, due to the magnetic sub-lattice site symmetry of Cr$_2$O$_3$, one has that $\mu_{\alpha\beta}^\mathrm{eff} = \mu^\uparrow_{\alpha\beta} = \mu_{\alpha\beta}^\downarrow$, see Supplementary material S2~\cite{MCA2026}.\\
\section{Results and Discussion}
\subsection{Intrinsic mobility} 
The crystal structure and the electron and hole mobility of Cr$_2$O$_3$ as a function of temperature are presented in Fig.~\ref{Cr2O3-mob}. 
Our results reveal a fundamental transport characteristic: the electron mobility is greater than the hole mobility at all temperatures, indicating that  Cr$_2$O$_3$ would be an intrinsic $n$-type semiconductor, a finding that provides a new perspective on the materials carrier transport properties.

To assess the reliability of the calculations, we compare the calculated hole mobilities with  the experimental values of the hole drift mobility at room temperature reported in Refs.~\cite{Singh2025,Qiu2024} as solid stars in Fig.~\ref{Cr2O3-mob}(b).
Comparing the calculated hole mobility (1-4 cm$^2$/(V$\cdot$s)) with the experimental values of the hole drift mobility ranging from 1.5~cm$^2$/(V$\cdot$s) to 11~cm$^2$/(V$\cdot$s), we find that the calculations agree rather well with experiments.
The observed experimental spread is largely attributed to the device architecture employed, organic field effect transistors (OFETs) and thin film transistors (TFT), where the gate and the electrode contact resistance can induce a much higher nominal mobility $\mu^\mathrm{nom}$ than the intrinsic one $\mu^\mathrm{int}$~\cite{Bittle2016,Hu2017}. 
For that reason, we also provide the saturation mobility $\mu^\mathrm{sat}$ of Ref.~\cite{Singh2025} (red star in Fig.~\ref{Cr2O3-mob}(b)), which following Ref.~\cite{Bittle2016} should be smaller than the one obtained in the linear region (green and yellow stars in Fig.~\ref{Cr2O3-mob}(b)). 
This agreement is further supported by comparing the temperature dependence of the hole drift mobility with the experimental effective Hall mobility that can be extracted from Ref.~\cite{Julkarnain2011}, see Supplementary Material S3~\cite{MCA2026}, confirming the robustness of the BTE approach.

This intrinsic $n$-type behavior stands in contrast with the common trend of $p$-type doping Cr$_2$O$_3$ of many studies, where it is assumed to be a $p$-type semiconductor \cite{Mohanapandian2021,Arca2017,Uekawa1996,Arca2011,Jagadish2024,Park1990,Jella2017,Crawford1964}. 
The origin of this discrepancy comes from the fact that Cr$_2$O$_3$ is a wide band gap semiconductor where experimental reports show an optical band gap of $3.1-3.3$~eV~\cite{Patel2023,Singh2019,Abdullah2014}. 
As a result, the intrinsic number of carriers will be relatively small, making it sensitive to defects, impurities, and oxidation~\cite{Holt1999}. 
Recent calculations by Wang and Mu~\cite{Wang2025} show that under oxygen-rich growth conditions, chromium vacancies (V$_\mathrm{Cr}$) exhibit exceptionally low formation energies, acting as acceptors and yielding the observed $p$-type character. Completing this picture, Medasani \textit{et al.}~\cite{Medasani2018,Medasani2019} made a similar analysis further showing that Cr-rich environments oxygen vacancies (V$_\mathrm{O}$) are dominant acting as donors ($n$-type doping). This aligns perfectly with our transport predictions. While the pristine lattice fundamentally favors electron mobility ($\mu_\mathrm{e}/\mu_\mathrm{h}=2-3$) due to its electronic structure, optimizing chemical synthesis towards oxygen reduced and Cr-rich environments favor donor defects to unlock this intrinsic $n$-type channel. This unified thermodynamic and carrier transport picture directly explains the experimental success in producing $n$-type Cr$_2$O$_3$ films via metallic Cr interfaces~\cite{Jella2017} or Ti-doping under reduced oxygen environments~\cite{Holt1999}. Thus, the systematically found $p$-type character of Cr$_2$O$_3$ is extrinsic. 
We verify that these trends are confirmed when using DFT+U, see Supplementary Material S4~\cite{MCA2026}.

Beyond its intrinsic mobility, Cr$_2$O$_3$ exhibits a conducting easy axis anisotropy along the $z$-axis yielding $\mu_{xx} = \mu_{yy} \neq \mu_{zz}$ in line with experimental reports of the conductivity~\cite{Crawford1964}. 
Comparing the in plane $\mu_{||} = \mu_{xx} = \mu_{yy}$ and out of plane $\mu_\perp = \mu_{zz}$ mobilities of both electrons and holes, we find that $\mu_{\perp} \sim 1.06\mu_{||}$, see Fig.~\ref{Cr2O3-mob}(b).
This anisotropy is particularly interesting because it would be aligned with its magnetic and magneto-electric (ME) easy axis.
Moreover, we note that although the easy axis anisotropy is qualitatively well captured by DFT calculations, using a DFT+U approach does not lead to better results, see Supplementary Material S4~\cite{MCA2026}. 

\subsection{Advantages of Cr$_2$O$_3$ as a TCO and beyond}
In the context  of TCO applications, Cr$_2$O$_3$ presents an electron mobility ranging between $\mu^\mathrm{e} \sim 5-10$~$\mathrm{cm^2/(Vs)}$ which is rather mediocre compared with other commonly used $n$-type TCOs such as ITO (In$_2$O$_3$ and SnO$_2$)  or ZnO , see Table~\ref{TCO-table}.  
In contrast, the hole mobility $\mu^\mathrm{h} \sim 1-4$~$\mathrm{cm^2/(Vs)}$ is suitable for TCO applications since commercially available $p$-type TCOs such as SnO, or CuAlO$_2$ 
show similar mobilities~\cite{Hautier2013,Singh2024,Willis2021}, see Table~\ref{TCO-table}. 
\begin{figure*}[t]
    \centering
    \includegraphics[width=0.8\textwidth]{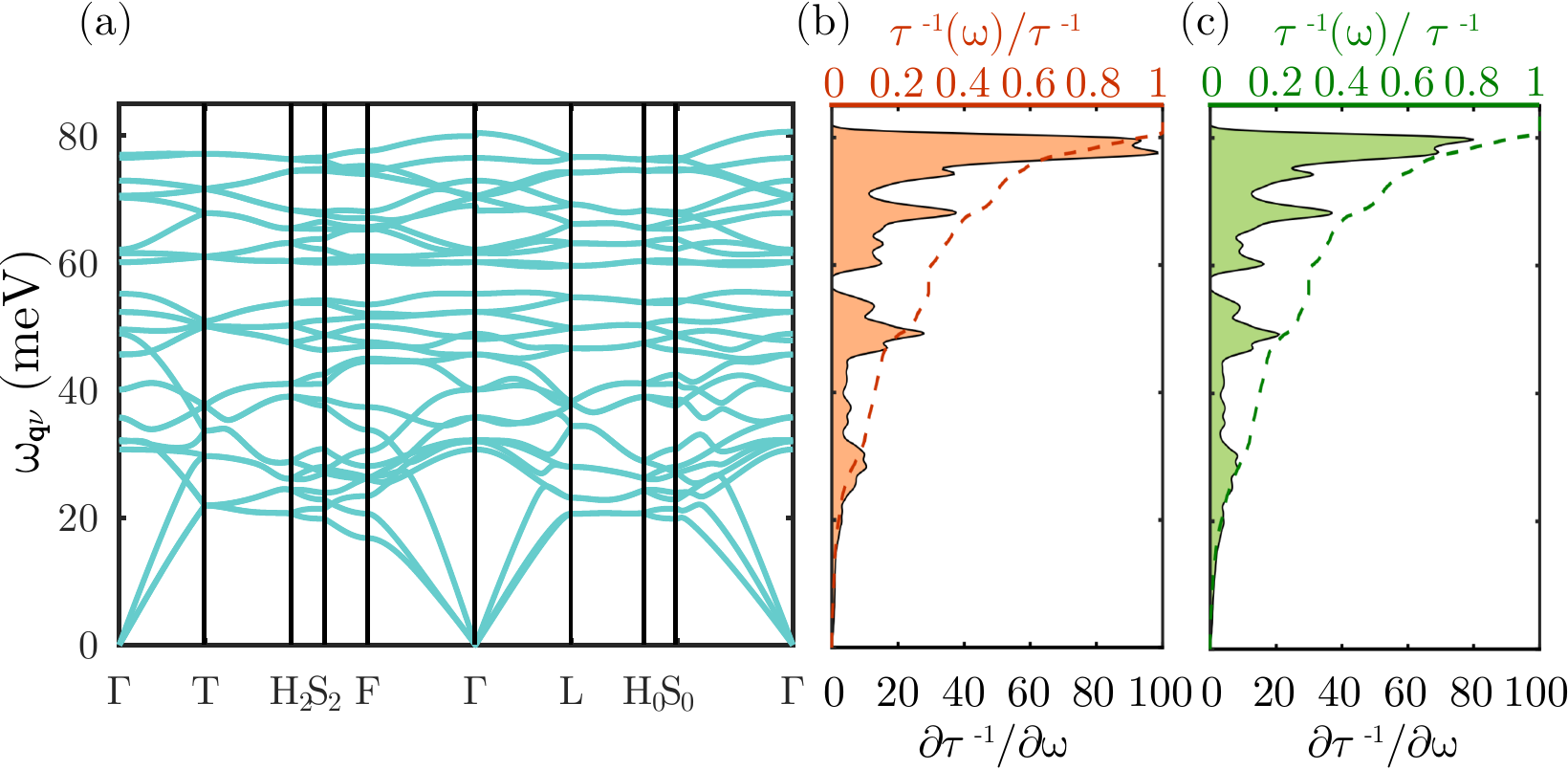}
    \caption{\label{Cr2O3-phon-spec}
    (a) Phonon dispersion of Cr$_2$O$_3$, and  spectral decomposition of the contribution of each phonon with energy $\omega_{\mathbf{q}\nu}$ to the total scattering rate at the band edges $\tau^{-1}$ for (b) hole mobility and (c) electron mobility. The total scattering rates are $\tau^{-1}_{\mathrm{CB}} = 851$~meV and $\tau^{-1}_{\mathrm{VB}} = 1020$~meV for electrons and holes, respectively. 
    }
\end{figure*}
Crucially, Cr$_2$O$_3$ exhibits a near parity electron and hole mobility ratio ratio $\mu^\mathrm{e}/\mu^\mathrm{h} = 2-3$, a feature that is strikingly absent in conventional high-performance TCOs, see Table~\ref{TCO-table}. While materials like In$_2$O$_3$ or ZnO offer higher absolute electron mobilities, their extreme carrier asymmetry (often exceeding 100:1) represents a fundamental bottleneck for complementary logic.
Compared to SnO -- a rare example of $p$-type oxide with a relatively balanced mobility ratio $\mu^\mathrm{h}/\mu^\mathrm{e} = 5-15$-- Cr$_2$O$_3$ offers wider optical window with a bandgap of 3.1-3.3~eV which ensures high transparency in the visible range, whereas SnO (2.7-3.0~eV) often suffers from a yellowish tint or absorption at the blue end of the spectrum. This makes Cr$_2$O$_3$ the more viable candidate for high-fidelity transparent displays and UV-photo-detectors.
Furthermore, Cr$_2$O$_3$ offers a distinct chemical processing advantage over low-dimensional or narrow gap alternatives. While $p$-type favored options like SnO are prone to phase structure instability and unintended oxidation into SnO$_2$ during high temperature processing, Cr$_2$O$_3$ is a thermodynamically stable oxide. This structural robustness and stability ensures that its balanced transport properties can be preserved during standard high-temperature oxide processing. 
While the absolute mobilities are not particularly high, Cr$_2$O$_3$ could be a good candidate for specialized targeted applications like monolithic, fully transparent optoelectronic circuits or ultraviolet (UV) photo-detectors requiring symmetric carrier extraction. Thus, using materials like Cr$_2$O$_3$ with sub-10 mobility ratios would challenge the current fast-electron/slow-hole paradigm of the TCO industry.\\
\indent In addition, the conducting and magnetic easy axis alignment in Cr$_2$O$_3$ provides a direct link between charge transport and magnetic order, suggesting that the carrier transport might be sensitive to the AFM state.
Combining this alignment of AFM Cr$_2$O$_3$ with its  large optical bandgap, reasonable $p$-type mobility and uniquely balanced electron and hole mobilities, places Cr$_2$O$_3$ as a multifunctional material allowing to build transparent, bipolar ME-driven AFM switching devices.

\subsection{Origin of the $n$-type nature of Cr$_2$O$_3$} In order to rationalize the difference between electron and hole mobilities, we compute the phonon dispersion and the frequency resolved scattering rate $\partial\tau^{-1}/\partial\omega$ for both electrons and holes. 
As seen in Fig.~\ref{Cr2O3-phon-spec}, most of the scattering for both electron and hole conduction come from optical phonons. 
We quantify the relative importance of each modes by computing the integrated scattering rate $\tau^{-1}(\omega)$. 
We find that for both electrons and holes, 90\% of the total scattering rate is given by optical phonons. 
Focusing on holes, we find that 50\% of the total scattering comes from the highest energy phonons at energies $75-82$~meV. 
The remaining 40\% is mainly given by slightly lower energy optical phonons of $67-70$~meV (25\% of the total scattering rate) and $47-51$~meV (10\% of the total scattering rate), see Fig.~\ref{Cr2O3-phon-spec}(a) and \ref{Cr2O3-phon-spec}(b). 
\begin{figure}[t]
    \centering
    \includegraphics[width=0.8\linewidth]{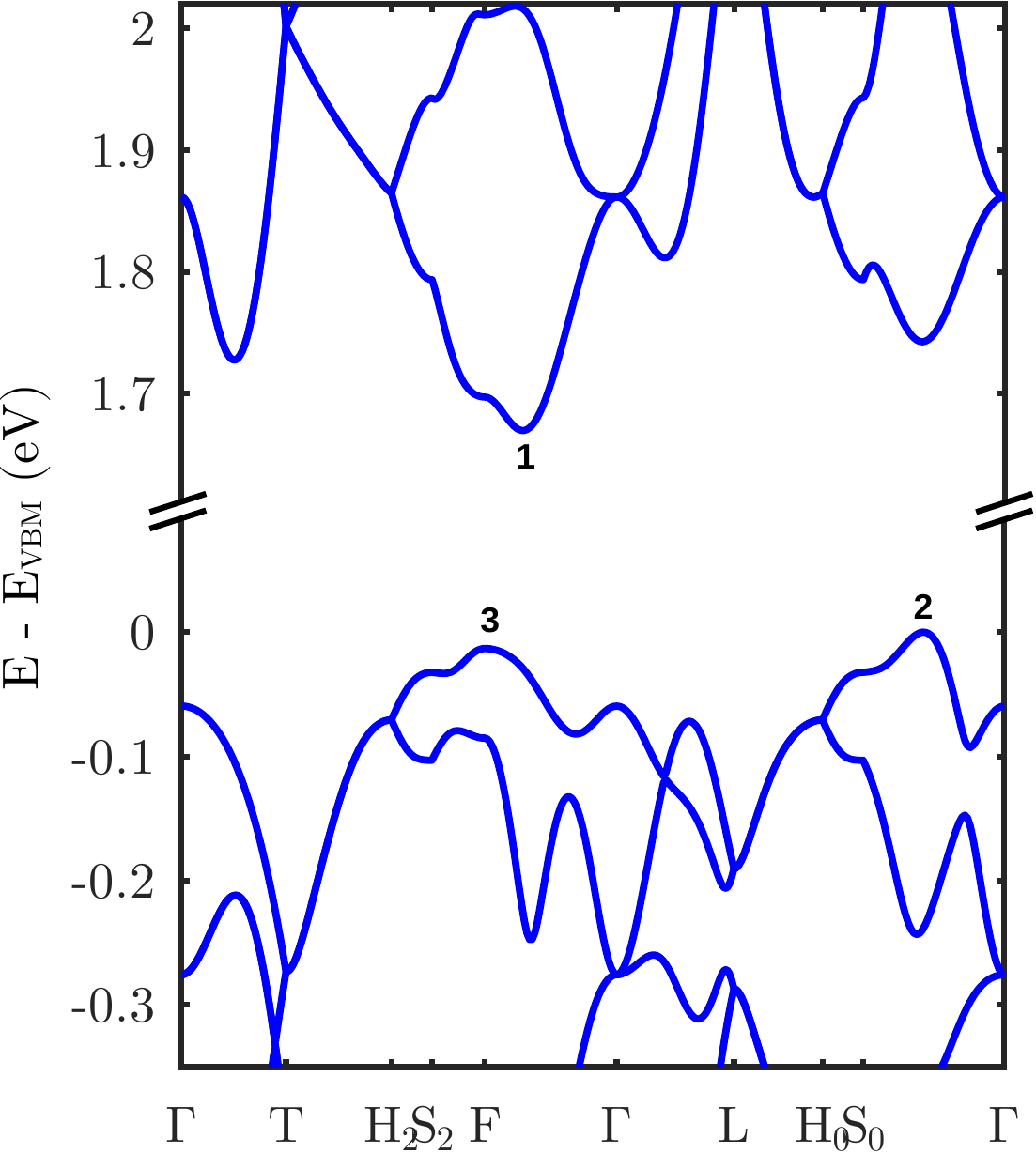}
    \caption{\label{Cr2O3-bands}
    Electronic dispersion of antiferromagnetic Cr$_2$O$_3$. 
    Conduction and valence band manifolds in the interesting region for mobility.
    }
\end{figure}

A similar analysis can be made in the case of the electrons where mainly optical phonons at similar energy scales produce the higher scattering rates with slightly less pronounced contributions from the phonons at energies $47-51$~meV than in the hole case, see Figs.~\ref{Cr2O3-phon-spec}(a) and \ref{Cr2O3-phon-spec}(c). 
In addition, the scattering rate distributions shown in Figs.~\ref{Cr2O3-phon-spec}(b) and \ref{Cr2O3-phon-spec}(c) are not only qualitatively but quantitatively similar. 
This implies that phonon scattering affects in a similar way both electrons and holes with only a 15$\%$ difference. 
As a consequence, the specific phonon modes that produce scattering cannot explain the disparity between the electron and hole mobility.

Therefore, we turn our attention to the electronic properties of Cr$_2$O$_3$. 
We compute the band structure and show in Fig.~\ref{Cr2O3-bands} the conduction and valence band edges relevant to the mobility in the given temperature range. 
We observe that the conduction band has 3 main valleys along the $\Gamma-\mathrm{T}$, $\Gamma-\mathrm{S}_0$ and $\Gamma-\mathrm{F}$, with the latter hosting the conduction band minimum (CBM), see point 1 in Fig.~\ref{Cr2O3-bands}. 
In contrast, the valence band has 4 valleys along the $\Gamma-\mathrm{T}$, $\Gamma-\mathrm{L}$, $\Gamma-\mathrm{S}_0$ and $\Gamma-\mathrm{F}$  paths, with the latter 2 being quasi-degenerate and hosting the VBM, see points 2 and 3 in Fig.~\ref{Cr2O3-bands}. 
This multi-valley structure of the valence band edge implies that for the holes the intra-valley scattering would be more pronounced than for the electrons.

Additionally, we note that the bands in the conduction band manifold are more dispersive than in the valence band. 
This indicates that in the conduction band we have higher band velocities and lower effective masses than in the valence. 
We compute the effective masses of the band edge of the valleys 1 and 2 in Fig.~\ref{Cr2O3-bands} that host the CBM and the VBM, respectively. 
We find that in the case of the electrons the effective mass is lighter than in the hole case. 
For the CBM $m^*\sim4m_e$ while for the VBM $m^* \sim 6m_e$. 
Thus, the main origin of the disparity is electronic with a minor role for the phonons. 
This can be further verified from the DFT+U calculations since the main difference are changes in band curvature and relative valley energy, see Supplementary Material S4~\cite{MCA2026} for further discussion.

\section{Conclusions} 
In summary, our \textit{ab initio} BTE calculations reveal that the phonon limited mobility of pristine Cr$_2$O$_3$
inherently favors $n$-type over $p$-type behavior, featuring a well balanced carrier mobility ratio $(\mu_\mathrm{e}/\mu_\mathrm{h} = 2-3)$. 
By contextualizing the intrinsic electron and hole mobility, its wide band gap, and the established thermodynamic and defect models, we solve a long-standing debate in the literature: the $p$-type character of Cr$_2$O$_3$ is an extrinsic feature dictated by oxygen rich  defect thermodynamics, whereas the pristine crystal is an intrinsic $n$-type semiconductor. Hence, the current trend of $p$-type doping Cr$_2$O$_3$ is only partially justified, since the hole mobility is around $\mu^\mathrm{h} \sim 1-4$~$\mathrm{cm^2/(Vs)}$, Cr$_2$O$_3$ and it could still be appealing in the context of $p$-type TCOs since they show lower mobility. Thus, the combination of a wide band gap, chemical and thermodynamic phase stability and  the near-parity electron-hole mobilities positions Cr$_2$O$_3$ as a unique platform for transparent bipolar electronics and functional oxide heterostuctures, shifting the design focus on oxygen reduction of Cr-rich processing regimes. In addition, we uncover the importance of the electronic degrees of freedom over the scattering with the phonons in Cr$_2$O$_3$ when determining charge carrier mobility. As a consequence, small changes in Cr$_2$O$_3$ electronic structure are expected to impact the mobility which hints at an exploitable response to external stimuli such as strain. Importantly, the alignment of the conducting easy axis of Cr$_2$O$_3$ with its known ME easy axis for both electron and hole mobility, suggests that Cr$_2$O$_3$ could be an interesting platform for transparent ME-FET and antiferromagnetic spintronic optoelectronic devices. Hence reshaping the role of Cr$_2$O$_3$ as a TCO and presenting itself as well as an interesting platform for bipolar transparent AFM spintronics, and overall as a multifunctional material.\\

\begin{acknowledgments}
\textit{Aknowledgments--}S. P. is a Research Associate of the Fonds de la Recherche Scientifique - FNRS.
This work was supported by the Fonds de la Recherche Scientifique - FNRS under Grants number T.0183.23 (PDR) and T.W011.23 (PDR-WEAVE). 
Computational resources have been provided by the supercomputing facilities of the Université catholique de Louvain (CISM/UCL) and the Consortium des Équipements de Calcul Intensif en Fédération Wallonie Bruxelles (CÉCI) funded by the Fond de la Recherche Scientifique de Belgique (F.R.S.-FNRS) under convention 2.5020.11 and by the Walloon Region. 
The present research benefited from additional computational resources made available on Lucia, the Tier-1 supercomputer of the Walloon Region, infrastructure funded by the Walloon Region under the grant agreement n°1910247. In addition, we acknowledge EuroCC Belgium for awarding this project access to the LUMI supercomputer, owned by the EuroHPC Joint Undertaking, hosted by CSC (Finland) and the LUMI consortium through EuroCC Belgium (UCLouvain-MODL-EPIMS-Bench-T0447)
\end{acknowledgments}

\bibliography{bibliography.bib}
\end{document}


\renewcommand{\thefigure}{S\arabic{figure}}
\renewcommand{\thetable}{S\arabic{table}}
\renewcommand{\thesection}{S\arabic{section}}
\newcommand\SP[1]{{\color{blue}[SP:#1]}}

\title{Supplementary Material: Challenging the $p$-type Paradigm: Intrinsic $n$-type Mobility in Antiferromagnetic Cr$_2$O$_3$}
\author{\'Alvaro Adri\'an Carrasco \'Alvarez}
\email{alvaro.carrasco@uclouvain.be}
\affiliation{European Theoretical Spectroscopy Facility, Institute of Condensed Matter and Nanosciences, Université catholique de Louvain, Chemin des Étoiles 8, B-1348 Louvain-la-Neuve, Belgium}
\author{Samuel Ponc\'e}%
\email{samuel.ponce@uclouvain.be}
\affiliation{European Theoretical Spectroscopy Facility, Institute of Condensed Matter and Nanosciences, Université catholique de Louvain, Chemin des Étoiles 8, B-1348 Louvain-la-Neuve, Belgium}
\affiliation{WEL Research Institute, avenue Pasteur 6, 1300 Wavre, Belgium.}
\date{\today}
\maketitle

\section{S1. Polaron radius}
In this section, we estimate if Cr$_2$O$_3$ is in the large or small polaron regime, following Ref.~\cite{Schlipf2018}. 
To that end, we compute the static dielectric constant and the Born effective charges as well as the effective mass of both electrons and holes and use it to make an estimate of the coupling strength of the Feynman polaron model $\alpha$ defined as:
\begin{equation}
    \alpha \equiv \frac{e^2}{4\pi\varepsilon_0\hbar} \sqrt{\frac{m^*}{2\hbar\omega_\mathrm{eff}}}\left(\frac{1}{\varepsilon(\infty)}-\frac{1}{\varepsilon(0)}\right)
\end{equation}
where $m^*$ is the effective mass, $\omega_\mathrm{eff}$ is the characteristic optical phonon frequency, $\varepsilon_0$ is the vacuum permittivity, $\varepsilon(\infty)$ is the high frequency dielectric constant (this is the dielectric constant $\varepsilon(\omega)$ when $\omega\rightarrow\infty$) and $\varepsilon(0)$ is the static dielectric constant,(this is the dielectric constant $\varepsilon(\omega)$ when $\omega = 0$) which we estimate as
\begin{equation}
    \varepsilon(0) = \varepsilon(\infty)+\frac{1}{V_{\mathrm{uc}}\varepsilon_0}\sum_\nu\frac{I_\nu}{\omega_{\nu}^2} 
\end{equation}
with $V_{\mathrm{uc}}$ being the volume of the unit cell, $I_\nu$ the infra red activities of the phonon mode with frequency $\omega_\nu$.
Regarding the effective phonon frequency $\omega_\mathrm{eff}$, we take  an average phonon frequency based on their infrared activity
\begin{equation}
    \omega_\mathrm{eff} = \frac{\sum_\nu \omega_\nu I_\nu}{\sum_\nu I_\nu}
\end{equation}
obtaining a value of $\omega_\mathrm{eff} = 63$~meV. 
Although the dielectric constant is a tensor and in Cr$_2$O$_3$ has 2 independent directions, we take an isotropic average for both static $\varepsilon$ and $\varepsilon_\infty$ to give an estimate of the coupling constant. 
For the effective mass, we consider the effective masses at the conduction band minimum (CBM) and valence band maximum (VBM). 
We obtain $m^*\sim4m_e$ for electrons and $m^* \sim 6m_e$ for holes which yields a coupling constant of $\alpha = 1.15$ for electrons and $\alpha=1.41$ for holes, which falls in the weak to intermediate coupling regime. 
In addition, we can estimate the polaron radius as $r_p = \sqrt{3.4\hbar/m^*\alpha\omega_\mathrm{eff}}$ obtaining values of 9.5~$\rm \AA$ and 7~$\rm \AA$ for electron and holes respectively, corresponding to 1.8 and 1.3 lattice parameters encompassing 23 and 9 unit cells respectively.
This falls in the large polaron regime following Ref.~\cite{Melo2023}.
Thus, we conclude that in Cr$_2$O$_3$, the transport properties can be well described within the Boltzmann transport equation since the polaron radius is in the large-intermediate regime, where band transport dominates over polaron hopping.

\section{S2. Spin resolved mobility}
\begin{figure*}[t]
    \centering
    \includegraphics[width=0.8\linewidth]{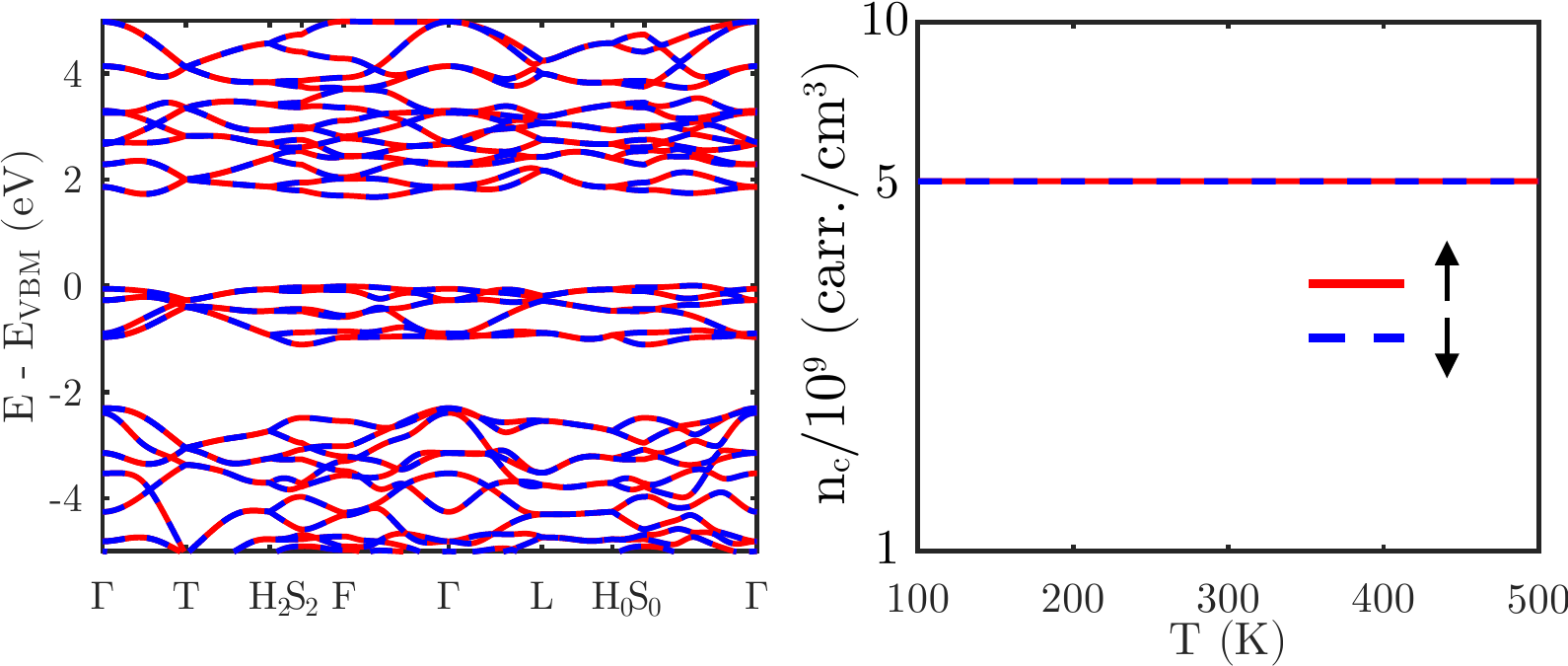}
    \caption{Electronic band dispersion for bulk Cr$_2$O$_3$ (a), and carrier density $n_c$ for different temperatures (b). 
    Red solid lines correspond to the spin up while dashed blue lines correspond to the spin down. 
    The total carrier density is set to $10^{10}$~cm$^{-3}$ at all temperatures.}
    \label{Cr2O3-bands-carrier-up-down}
\end{figure*}

In this section, we calculate the relative contributions of each spin channel in Cr$_2$O$_3$ showing that due to the lattice site symmetry the spin resolved mobilities are the same $\mu_{\alpha\beta}^\uparrow = \mu_{\alpha\beta}^\downarrow$ as well as the carrier density $n^\uparrow_c = n^\downarrow_c$. 
We first compute the band structure for the two spin channels in Fig.~\ref{Cr2O3-bands-carrier-up-down}(a) which is a hallmark of the lattice site symmetry. 
Additionally, we compute the spin resolved mobility and the relative carrier density as a function of temperature in Fig.~\ref{Cr2O3-bands-carrier-up-down}(b).

\begin{figure*}[t]
    \centering
    \includegraphics[width=0.7\linewidth]{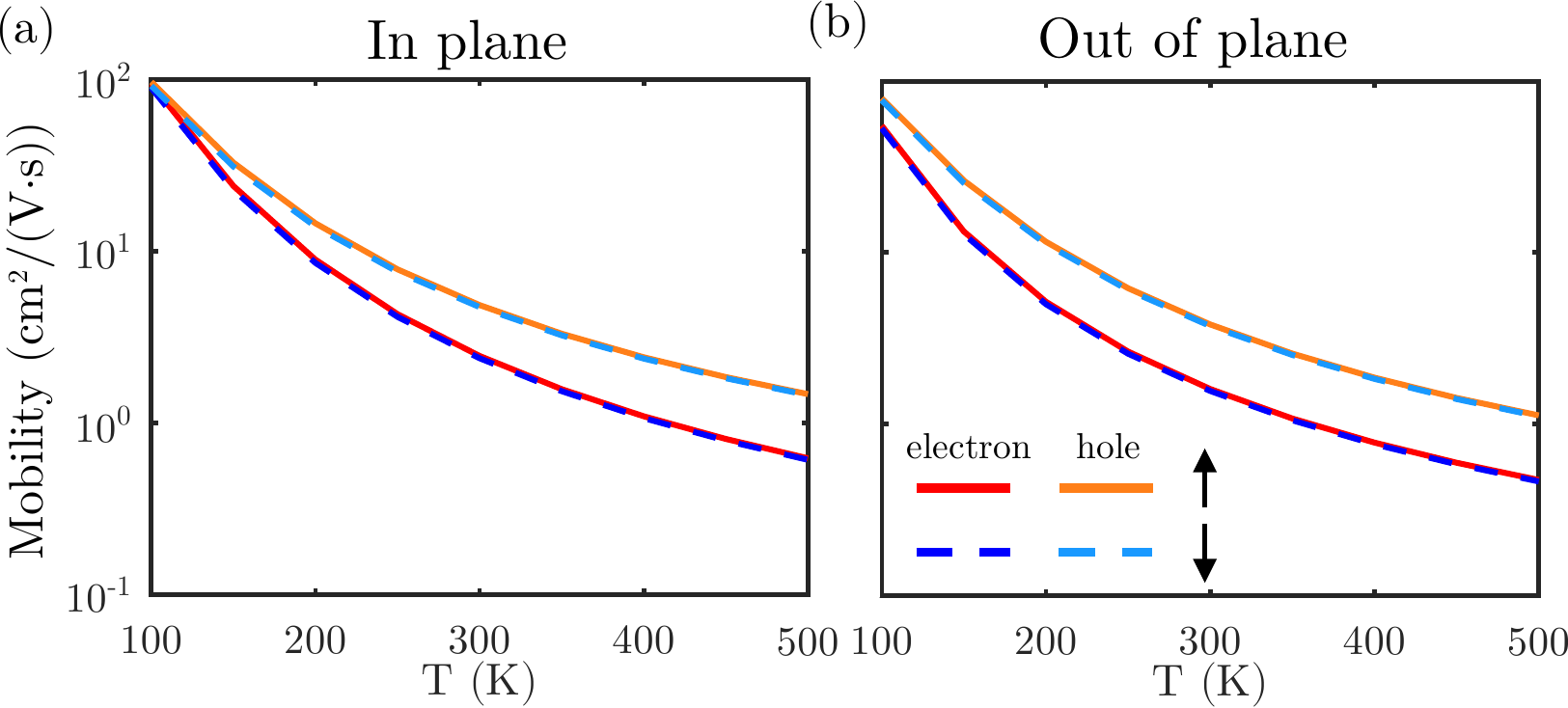}
    \caption{Electron and hole mobility for bulk Cr$_2$O$_3$. (a) In plane $\mu_{||} = \mu_{xx} = \mu_{yy}$ and (b) out of plane $\mu_{\perp} = \mu_{zz}$. 
    Solid lines correspond to the spin up (red and orange for electron and hole, respectively) while dashed lines correspond to the spin down (dark and light blue for electron and hole mobility, respectively). 
    The total carrier density is set to $10^{10}$~cm$^{-3}$ at all temperatures.}
    \label{Cr2O3-mob-up-down}
\end{figure*}

We see that the eigenvalues are spin degenerate $\varepsilon_{n\mathbf{k}}^\uparrow=\varepsilon_{n\mathbf{k}}^\downarrow$.
This yields an equivalent relative carrier density between the two spin channels $n_c^\uparrow=n_c^\downarrow$. 
This is not the case in general even if the system is antiferromagnetic. 
The combination of these two factors yields mobilities that are equivalent as well $\mu_{\alpha\beta}^\uparrow = \mu_{\alpha\beta}^\downarrow$, see Figs.~\ref{Cr2O3-mob-up-down}(a) and \ref{Cr2O3-mob-up-down}(b) for the in plane and out of plane electron and hole mobilities, respectively.

\newpage
\clearpage

\section{S3. Temperature dependence of the Hall mobility}
\begin{figure}[h]
    \centering
    \includegraphics[width=0.4\linewidth]{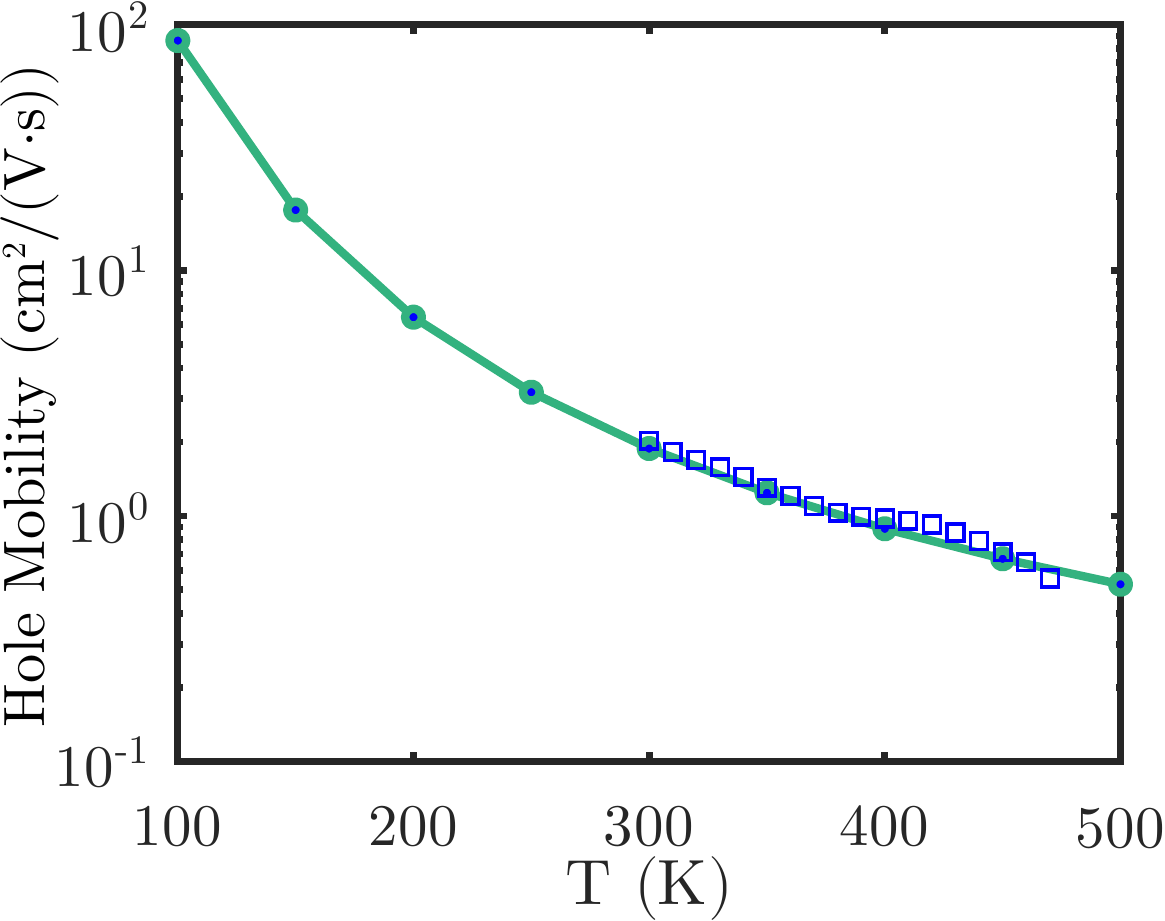}
    \caption{Hole mobility of antiferromagnetic Cr$_2$O$_3$. Blue squares correspond to the Hall mobility computed as $\mu_\mathrm{H} = \sigma R_\mathrm{H}$ and scaled by a factor $\frac{1}{4}$. Green solid line correspond to the calculated hole drift mobility.}
    \label{Cr2O3-mob-Hall}
\end{figure}

We compare the hole Hall mobility from the experimental data of Ref.~\cite{Julkarnain2011}, with the drift mobility obtained with the Boltzmann Transport Equation, see Fig.~\ref{Cr2O3-mob-Hall}. 

The experimental Hall mobility $\mu_\mathrm{H} = \sigma R_\mathrm{H}$ is obtained as the product of the Hall resistance $R_\mathrm{H}$ and the conductivity $\sigma$. 
Additionally, it is common for the Hall mobility to be higher than the drift mobility since the Hall factor $\mu_\mathrm{H}/\mu = r_\mathrm{H}$ is often greater than one. 
Although strictly speaking, the Hall factor is temperature dependent, its variation is not extremely pronounced~\cite{Ponce2021} and taking it to be constant is reasonable approximation. 
In that regard, we scale the experimental Hall values computed as $\mu_\mathrm{H}=\sigma R_\mathrm{H}$ Ref.~\cite{Julkarnain2011} by a constant factor of $\frac{1}{4}$, see blue squares in Fig.~\ref{Cr2O3-mob-Hall}, to better compare the temperature dependence. 
In doing so, we find that the temperature dependence is well described by our calculations.

\section{S4. Mobility calculations with DFT+U}
\begin{figure}[b]
    \centering
    \includegraphics[width=0.4\linewidth]{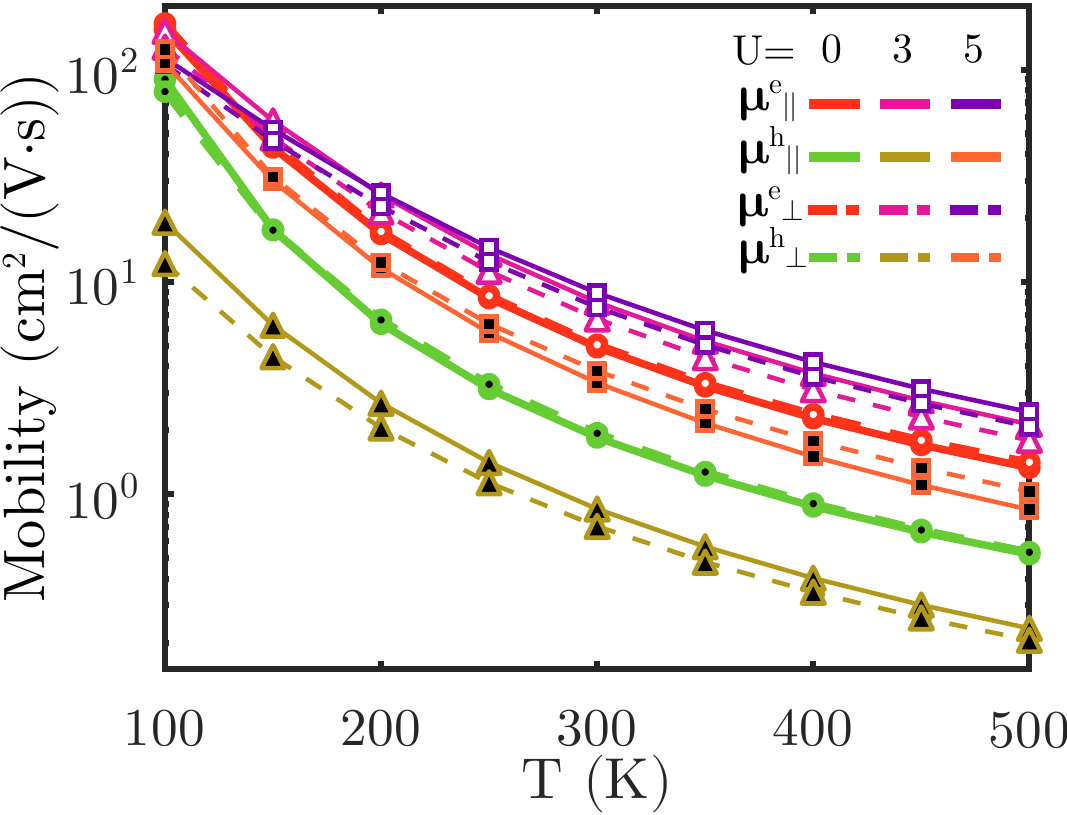}
    \caption{\label{Cr2O3-mob-DFTU}
    Intrinsic electron (red, magenta and purple lines) and hole (green, khaki and orange lines) mobility for bulk Cr$_2$O$_3$ as a function of temperature obtained with the Boltzmann transport equation (BTE). Dashed lines correspond to out of plane mobility while solid lines are in plane mobility values. A DFT+U approach is taken with $U = 0,3,5$~eV corresponding to the curves with circles, triangles and squares respectively.}
\end{figure}

\begin{figure*}[t]
    \centering
    \includegraphics[width=0.8\linewidth]{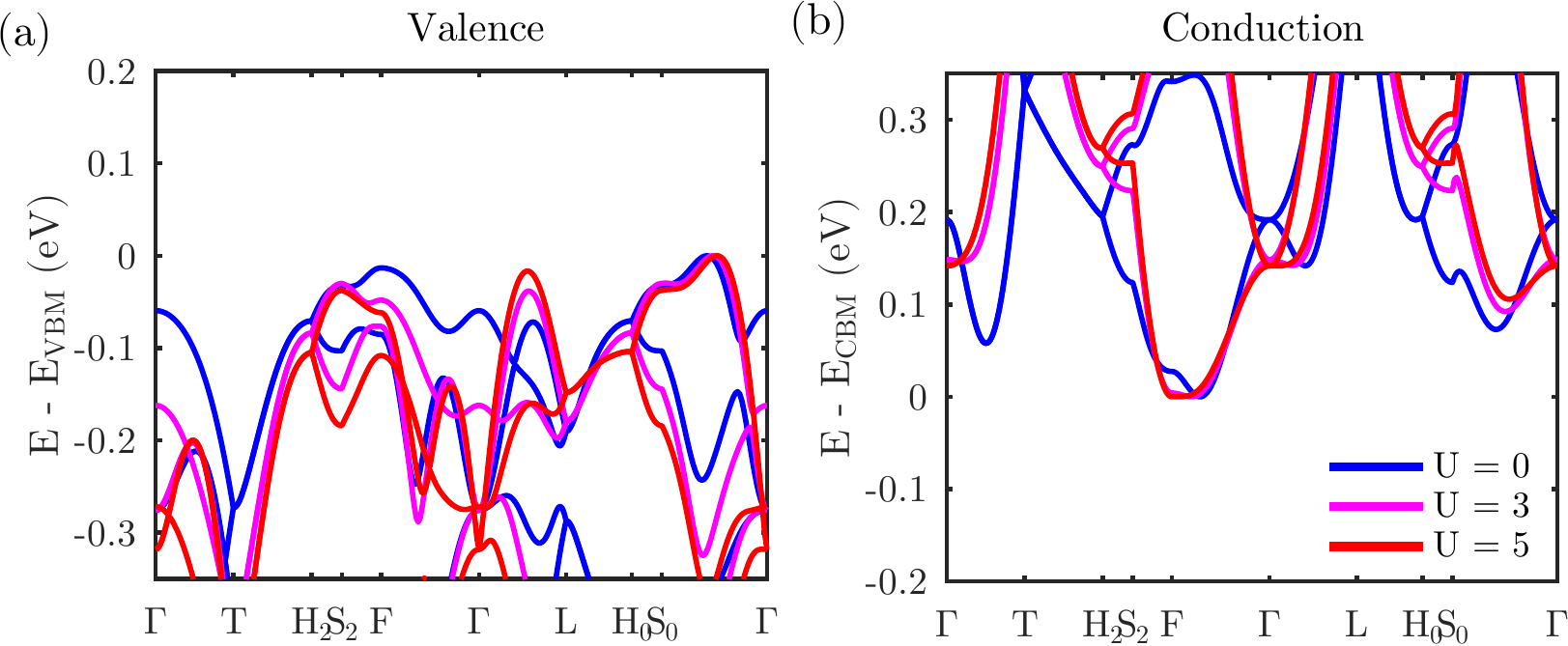}
    \caption{Valence band (a) and conduction band (b) manifolds of Cr$_2$O$_3$. A DFT+U approach is used with $U = 0,3,5$ eV corresponding to blue, magenta and red bands.}
    \label{Cr2O3-bands-DFTU}
\end{figure*}

In this section we show the mobility calculations using a DFT+U ~\cite{Yang2025} for both electron and hole mobility considering the three cases $U = 0,3,5 $~eV. 
The results are given in Fig.~\ref{Cr2O3-mob-DFTU} for relaxed lattice parameters for each $U$ value while keeping the same $\mathbf{k}/\mathbf{q}$ meshes as with DFT. 
The general trends with respect to the $n$-type behavior of Cr$_2$O$_3$ are not affected by the specific $U$, giving the same qualitative picture. 
Nonetheless, we notice that at the quantitative level, the inclusion of the onsite $U$ repulsion produces different behaviors for the electron and hole mobilities. 
While for the hole mobility on site $U$ can both enhance or reduce the hole mobility, for the electron mobility there is a consistent enhancement regardless of the $U$ value. 
This is mainly due to the change in band topology of valence and conduction band manifolds, see Fig.~\ref{Cr2O3-bands-DFTU}.

By comparing the different bands in the Fig.~\ref{Cr2O3-bands-DFTU}(a), it becomes clear that with DFT+U the inter-valley scattering is strongly affected by the specific $U$ values. 
The inclusion of non zero $U$ produces the disappearance of the $\Gamma-\mathrm{T}$ and $\Gamma-\mathrm{F}$ valleys when increasing the $U$, and at the same time the appearance of two valleys along  the $\mathrm{H}_2-\mathrm{F}$ direction and more importantly $\Gamma-\mathrm{L}$ directions which for larger $U$ reaches almost the Fermi level. 
For the electron mobility, the inter-valley scattering is reduced for non-zero $U$ values because the valleys along the $\Gamma-\mathrm{T}$ and $\Gamma-\mathrm{S}_0$ directions are pushed upwards in energy, see Fig.~\ref{Cr2O3-bands-DFTU}(b). 
In addition, we can find that in Fig.~\ref{Cr2O3-mob-DFTU} the conducting easy axis anisotropy disappears with a finite $U$ for the electron mobility while for the hole mobility the specific $U$ may change from easy to hard axis with larger $U$ values leading to a more prominent easy axis similar to experiments~\cite{Crawford1964}. 
For the case of the electrons, we can understand the change from easy axis to hard axis anisotropy by increasing U as a consequence of the tendency of DFT+U approaches to produce a much smaller band width, flatter bands and larger effective masses~\cite{Falletta2022}. 
We can verify this in  Figs.~\ref{Cr2O3-bands-DFTU}(a) and \ref{Cr2O3-bands-DFTU}(b) where it is clear that along the out of plane $\Gamma-\mathrm{T}$ and $\Gamma-\mathrm{F}$ directions the bands are flatter and less dispersive with non zero $U$ when compared with the in plane $\Gamma-\mathrm{L}$ and $\Gamma-\mathrm{S}_0$ directions. 
A possible approach to fix the issue of the $U$ values is to compute the self consistent $U$ following Refs.~\cite{Timrov2021}. 
We proceed with that approach and obtain a self consistent value of $U_\mathrm{sc} = 2.83\pm0.02$~eV which is quite similar to the $U = 3$~eV used here. 
Thus, although DFT+U approaches are successful when providing an effective tool to open band gaps, the band curvature is not always accurately predicted and in some cases can produce the wrong anisotropic behavior~\cite{Falletta2022}.

\bibliography{bibliography-SI.bib}